\documentclass[11pt]{article}
\usepackage{amsfonts}
\usepackage{amsmath,amssymb}
\usepackage{bbm}
\usepackage{bm}
\usepackage{color}
\usepackage{slashed}
\usepackage{cite}
\usepackage[utf8]{inputenc}
\usepackage{graphicx}
\usepackage{bbold}
\DeclareGraphicsExtensions{.pdf,.png,.jpg}
\usepackage[colorlinks=false,citecolor=cyan,urlcolor=blue,bookmarks=true,bookmarks=true,bookmarksopen=true,bookmarksnumbered=true,bookmarksopenlevel=3]{hyperref}

\definecolor{airforceblue}{rgb}{0.36, 0.54, 0.66}
\definecolor{steelblue}{rgb}{0.27, 0.51, 0.71}
\definecolor{amber}{rgb}{1.0, 0.49, 0.0}

\allowdisplaybreaks[1]
\def\simg{{\ \lower-1.2pt\vbox{\hbox{\rlap{$>$}\lower6pt\vbox{\hbox{$\sim$}}}}\ }}
\def\siml{{\ \lower-1.2pt\vbox{\hbox{\rlap{$<$}\lower6pt\vbox{\hbox{$\sim$}}}}\ }}

\makeatletter \@addtoreset{equation}{section} \makeatother

\begin{document}

\flushbottom

\begin{titlepage}

\begin{centering}

\vfill

{\Large{\bf
  Bound-state effects for dark matter with Higgs-like mediators
}} 

\vspace{0.8cm}

S.~Biondini %\footnote{biondini@itp.unibe.ch}

\vspace{0.8cm}

% $^\rmi{b}$%
{\em AEC, Institute for Theoretical Physics, 
University of Bern, \\ 
Sidlerstrasse 5, CH-3012 Bern, Switzerland} 

\vspace*{0.8cm}

\end{centering}

\vspace*{0.3cm}
 
\noindent
\textbf{Abstract}: %One of the most appealing hypothesis for the dark matter is that of a weakly interacting massive particle. In this scenario, complementary experimental searches constrain the parameter space of a given model, either full or some simplified version of it. %The relic abundance plays a key role to establish if the dark matter model is viable to
%Simplified models have been put forward to effectively classify dark matter candidates and set constrains on the parameter space from complementary experimental searches.   
%In many realistic models, the dark matter particle is part of a dark sector. Therefore, a more accurate derivation of the annihilation cross section that fixes the dark matter energy density is required: Sommerfeld enhancements, bound-state effects, thermal masses and widths can be relevant to the particle dynamics in the early universe. 
In this paper we study the impact of a scalar exchange on the dark matter relic abundance by solving a plasma-modified Schr\"odinger equation.  A simplified model is considered where a Majorana dark matter fermion is embedded in a U(1)$'$ extension of the Standard Model and couples with a dark Higgs via a Yukawa interaction. We find that the dark-Higgs exchange can increase the overclosure bounds significantly. For the largest (smallest) value of the Yukawa coupling examined in this work, the dark matter mass is lifted from 5 TeV (0.55 TeV) to 27 TeV (0.70 TeV). 

\vfill

\vfill
%\newpage
%\tableofcontents
\end{titlepage}

\section{Introduction}
\setcounter{page}{1}
Dark matter is one of the main open problems in the realm of cosmology and particle physics. If dark matter is assumed to be a particle rather than an astrophysical object, the hypothesis of a weakly interacting massive particle (WIMP) has been certainly the most studied. This choice does not fix a unique candidate though, on the contrary a plethora of possible dark matter particles are available\cite{Steffen:2008qp,Roszkowski:2017nbc}.  
The quest for a successful candidate poses interesting connections between the machinery of
quantum field theory, needed to calculate dark matter annihilation and scattering rates, and the many constraints imposed from the astrophysical and Earth-based experimental measurements. This has resulted in highly constrained scenarios: the viable parameter space of a given model is often in tension with that needed to reproduce the observed dark matter relic abundance via the so called freeze-out mechanism (see e.g.\ ref.\cite{Arcadi:2017kky} for a comprehensive status on WIMPs). Here, the key ingredient is the annihilation cross section of dark matter pairs that enters a Boltzmann equation and eventually determines the freeze-out abundance\cite{Lee:1977ua,Gondolo:1990dk,Griest:1990kh}. The latter has to match with the accurate measurement of the dark-matter energy density $\Omega_{\hbox{\tiny DM}} h^2= 0.1186 \pm 0.0020$ \cite{Ade:2015xua}. 

Recently, simplified models have been suggested for the
interpretation of beyond the Standard Model searches at colliders, direct and indirect detection experiments \cite{Abdallah:2015ter,DeSimone:2016fbz,Morgante:2018tiq}. In this framework, rather than considering
a fully fledged theory, bounds and constraints are set on a simple model that captures the most relevant physics.
Reinterpreting the experimental results in terms of simplified models, strong lower bounds are currently being set
by recent analyses at the Large Hadron Collider (LHC) \cite{Kahlhoefer:2017dnp} and the XENON1T experiment \cite{Aprile:2017aty,Aprile:2017iyp} that look for the footprint
of a new massive particle. Within simplified models, one is able to classify in a systematic way the nature of new degrees of freedom that may play the role of a dark matter particle, together with accompanying particles of the new physics model. Indeed in many cases, the so-called mediators act as  \textit{portals} between the dark and visible sector (it is also possible to have more than one mediator), preserve unitarity and gauge invariance, and enrich the phenomenology.  

When moving to such realistic particle models, some processes may occur that call for revisiting the standard relic abundance calculation, i.e.\ the derivation of the annihilation cross section in the early universe. For example, potential-like interactions are induced by a sufficiently light vector or scalar mediator (lighter than the dark matter mass) together with
the possibility of bound-state formation. For a mediator mass comparable with the dark matter mass, coannihilations can play an important
role and the mediator can itself experience soft interactions if coupled with light Standard Model degrees of freedom. Thermal masses and thermal interaction rates may also be important, the latter can lead to bound-state formation/dissociation in a thermal bath. The inclusion of some of these effects has led to  substantial revision of the overclosure bound for a given dark matter model, namely the largest value of the particle mass compatible with the observed dark matter energy density. In particular, the electroweak gauge boson exchange and gluon exchange can be important and the corresponding Sommerfeld enhancement has been included in the annihilation cross section in many studies, e.g. \cite{Feng:2010zp,Cirelli:2009uv, deSimone:2014pda, Harz:2017dlj}. The inclusion of bound-state effects in the annihilation process through a Boltzmann equation is rather non-trivial and different approaches have been put forward lately\cite{Detmold:2014qqa,vonHarling:2014kha,Liew:2016hqo,Kim:2016kxt,Mitridate:2017izz,Keung:2017kot,Biondini:2017ufr,Biondini:2018pwp}. 

A non-perturbative formalism for addressing
the thermal annihilation of non-relativistic particles has been developed quite recently \cite{Kim:2016zyy,Kim:2016kxt}. 
In this context, the thermally averaged annihilation cross section is obtained 
in terms of a chemical equilibration rate~\cite{Bodeker:2012gs}, the latter extracted from correlators evaluated
in equilibrium and independent of the assumptions typical of a Boltzmann description. The key ingredient is the
imaginary part of a two-point Green’s function, namely a spectral function. The advantage of using such an approach
is twofold: (i) the spectral function can be determined by solving a thermally-modified Schrödinger equation with
static potentials that comprise several in-medium effects like virtual and real scatterings; (ii) the appearance of bound
states is naturally described in this framework and the need of complicated bound-state production and dissociation
rates is avoided. This formalism has been applied to the Inert Doublet Model and to a simplified model comprising a Majorana fermion coannihilating with a strongly interacting scalar, where weakly and strongly bound states appear respectively \cite{Biondini:2017ufr,Biondini:2018pwp}.  

Potential-like interactions arise naturally when considering a fermion or a scalar dark matter coupled to gauge bosons (due to the trilinear vertex in the covariant derivative). However, it is also possible to have a scalar exchange between dark matter pairs, such as the Standard Model Higgs boson or the corresponding Higgs boson of the new physics model. In the latter case, we refer to it as dark Higgs throughout the paper. The effect of the Higgs boson exchange  has been studied for the Inert Doublet Model  with a focus on dark-matter annihilations leading to gamma ray signals \cite{Garcia-Cely:2015khw}, together with an estimate of the impact on cross sections in the early universe.  Similar analyses have been carried out for scalar and fermionic dark matter with a Higgs portal\cite{MarchRussell:2008yu,MarchRussell:2008tu,LopezHonorez:2012kv,Harz:2017dlj}. In all cases, the Sommerfeld effect has been studied that affects the dark matter pair wave function at zero temperature. In this work, we aim to apply the aforementioned finite-temperature formalism \cite{Kim:2016zyy,Kim:2016kxt} to assess the formation of bound state induced by a scalar exchange besides the Sommerfeld enhancement. We shall work in the framework of simplified models. The bulk of the analysis is carried out for a model with a spontaneously broken U(1)$'$ gauge symmetry that contains a Majorana dark matter fermion, a dark gauge boson and a dark Higgs\cite{Holdom:1985ag,Babu:1997st,Kahlhoefer:2015bea,
Duerr:2016tmh}. In addition, we elaborate on an another model of recent interest, namely a Majorana dark matter coannihilating with a coloured scalar charged under QCD and interacting with the SM Higgs boson \cite{Garny:2015wea}. 

The plan of the paper is the following. In section~\ref{sec_U1}, we discuss the simplified model that we focus on, i.e. a U(1)$'$ extension of the SM. In section~\ref{sec_U1_2_0} the thermally averaged annihilation cross section is presented within an effective field theory approach.  Then we derive the non-relativistic Lagrangian in section~\ref{sec_U1_2}, the thermal potentials are given in section~\ref{sec_U1_3}, whereas the plasma-modified Schr\"odinger equation is discussed in section~\ref{sec_U1_3} together with numerical outputs for the overclsoure bound. We consider other simplified models where a Higgs exchange can appear in section~\ref{sec_others}. Finally some conclusions and discussion are offered in section~\ref{sec_concl}.

\section{Majorana fermion dark matter and U(1)$'$ gauge symmetry}
\label{sec_U1_main}
We want to study dark matter models where a scalar field can be exchanged between the dark matter particles. As a well-motivated and interesting example, we pick the simplified model recently described in refs.\cite{Kahlhoefer:2015bea,
Duerr:2016tmh} that realizes perturbativity and gauge invariance at the same time. 
\subsection{Model description and light-mediators regime}
\label{sec_U1}
The model contains a dark Higgs and a dark gauge boson in addition to a Majorana fermion dark matter  (the latter is assumed to be the actual dark matter particle that contributes to the present universe energy density). The dark Higgs provides the mass of both the dark matter fermion and the dark gauge boson via the spontaneous breaking of the U(1)$'$ symmetry. Portal couplings induce an interaction between the dark and the SM sector (scalar mixing and gauge boson mixing).  The Lagrangian of the model reads~\cite{Kahlhoefer:2015bea,
Duerr:2016tmh}
\begin{eqnarray}
\mathcal{L}&=&\mathcal{L}_{\hbox{\tiny SM}}+\frac{1}{2}  \bar{\chi} \left( i \slashed{\partial}  -e' q_\chi  \gamma^5 \slashed{V}^\mu \right)  \chi-\frac{1}{2} y_\chi \bar{\chi} ( S P_L + S^*P_R ) \chi \nonumber \\
&+& (D^\mu S)^* (D_\mu S) + \mu_s^2 S^*S -\lambda_s (S^*S)^2  -\lambda_{hs} S^*S H^\dagger H \nonumber \\
&-&\frac{1}{4} V^{\mu \nu} V_{\mu \nu} -\kappa V^{\mu \nu} F_{\mu \nu} - e' V^\mu \sum q_f \bar{f} \gamma^\mu f  \, ,
\label{Lag_UV_u1}
\end{eqnarray}
where $\chi$ is a Majorana fermion field, $V^{\mu}$ is the dark gauge boson, $S$ is the dark Higgs field, $H$ is the Standard Model Higgs doublet, $V^{\mu \nu}= \partial^{\mu} V^{\nu} - \partial^{\nu} V^{\mu} $ and $F^{\mu \nu}= \partial^{\mu} B^{\nu} - \partial^{\nu} B^{\mu} $ where $B^{\mu}$ is the Standard Model U(1)$_Y$ gauge field. Then $f$ is a generic Standard Model fermion that couples via a vector current with the U(1)$'$ boson and $q_f$ is the corresponding charge. The fermion dark matter couples to the dark gauge boson with an axial-vector current (the vector current vanishes for a Majorana fermion and this choice helps in suppressing direct detection cross section with respect to the Dirac case). The covariant derivative acting on the dark Higgs field reads 
$D_\mu = \partial_\mu + i e'q_s V_\mu$.
In order to write the gauge invariant mass term for the Majorana dark matter in the first line of eq.~(\ref{Lag_UV_u1}), we have to require $ q_s=-2 q_\chi$\cite{Kahlhoefer:2015bea}. Then we define $g_\chi \equiv e' q_\chi$ and therefore $D_\mu = \partial_\mu - 2 i  g_\chi V_\mu$. In the following we neglect the portal couplings $\lambda_{hs}$ and $\kappa$.
\begin{figure}[t!]
\centering
\includegraphics[scale=0.47]{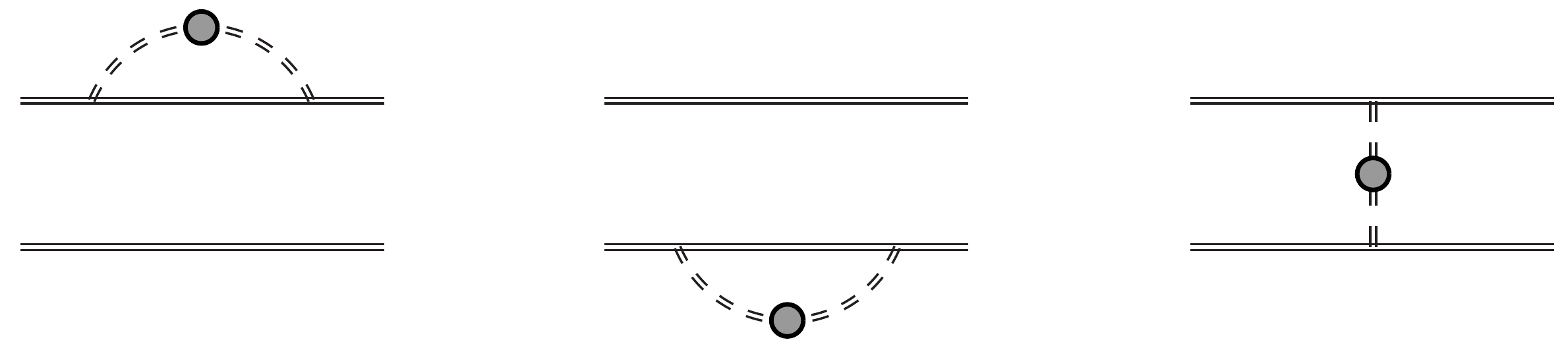}
\caption{\label{fig:Higgs_potential_1} Diagrams leading to a mass correction and an attractive potential between the dark matter fermion pair (double-solid line) induced by the light dark-Higgs scalar (double-dashed line). The blob stands for thermal correction to the scalar mass.}
\end{figure}

An important observation is that the couplings between the dark matter and the dark Higgs and the dark matter
and the dark gauge bosons are not independent \cite{Kahlhoefer:2015bea,
Duerr:2016tmh}. Indeed, after the U(1)$'$ symmetry breaking, $S=(w +s +i \varphi)/\sqrt{2}$, the two masses read in the $T=0$ limit (in general $w$ depends on the temperature, see section~\ref{sec_U1_3})
\begin{equation}
M_\chi= \frac{y_\chi w}{\sqrt{2}} \, , \quad m_V= 2 g_\chi w \, ,
\end{equation} 
and they are related to each other as 
\begin{equation}
\frac{M_\chi}{m_V}=\frac{y_\chi}{2 \sqrt{2} g_\chi} \, .
\label{eq_mass_break}
\end{equation}
According to the global analysis given in \cite{Duerr:2016tmh}, the model is rather unconstrained by experiments in the region where $M_\chi > m_V,m_s$. 
This is also the situation where one expects the dark-Higgs and dark-vector exchange to have some impact. Moreover, from eq.~(\ref{eq_mass_break}), one can see that requiring $M_\chi \gg m_V$ implies $y_\chi \gg g_\chi$. 
This suggests that the coupling between the dark Higgs and the dark matter is larger than the one between the dark matter and the dark gauge boson. % and the dark matter in the regime of light mediators (where light always refers to the comparison with the dark matter mass $M_\chi$). 
This is a hint to motivate the inspection of dark-Higgs exchange diagram, see figure~\ref{fig:Higgs_potential_1}. Furthermore, we also ask the dark matter to be heavier than the scalar mass. We can use the relation in the $T=0$ limit\footnote{We checked that at finite temperature the ratio changes by at most of  10\% at the freeze-out temperature, e.g.\ at  $T \simeq M_\chi /20$. Even if we include thermal masses for the dark Higgs in the following numerical study, we use the $T=0$ ratio $M/m_s$ to identify points in the parameter scan. The dark fermion mass is always taken in its $T=0$ limit.}
\begin{equation} 
M_\chi=\frac{y_\chi w}{\sqrt{2}} = \frac{y_\chi m_s}{2 \sqrt{\lambda_s}} \Rightarrow \frac{M_\chi}{m_s} = \frac{y_\chi}{2\sqrt{\lambda_s}} \, ,
\end{equation} 
and then pick the appropriate values for the couplings to fix the desired ratio $M_\chi/m_s \gg 1$.

Let us stress that, in this particular model, the dark matter mass is provided by the spontaneous breaking of the U(1)$'$ gauge symmetry. Therefore, only the broken phase is relevant to us in order to study the freeze-out mechanism: the dark matter has to acquire a finite mass $M_\chi$, attain thermal equilibrium and enter a non-relativistic regime when its mass drops below the plasma temperature. Eventually it decouples around $T \sim M_\chi/25...M_\chi/20$ like in the standard WIMP scenario. However, we notice that in the case $\lambda_{hs} \neq 0$  the dark and Standard Model Higgs expectation values are coupled and their evolution with temperatures may not be trivial (see appendix in ref.\cite{Kahlhoefer:2015bea} for more details on the scalar mixing).

\subsection{Dark-matter annihilations in a thermal bath}
\label{sec_U1_2_0}
Our aim is to describe accurately dark matter pair annihilations and include systematically near-threshold effects in a finite temperature environment, most importantly bound-state formation. Soft exchange processes are mediated by the dark Higgs and the gauge boson. %The particle exchanged by the dark matter fermions are the dark Higgs and the dark gauge boson.

First, let us summarize the framework of the freeze-out of a heavy thermal relic that puts us in a deep non-relativistic regime. The dark matter particles are kept in chemical equilibrium through 
interactions with the thermal bath until $T \ll M_\chi \equiv M$ and gradually freeze out at temperatures  
$T \sim M/25$. Annihilations continue even during later stages where the dark matter particles are still in kinetic equilibrium. In this situation most of the energy of a dark matter particle is given by its mass and, for non-relativistic species, the typical momentum is $|\bm{p}|=\sqrt{MT}= M \sqrt{T/M}$. One usually identifies an average velocity $v \equiv \sqrt{T/M}$, which is smaller than unity in the regime of interest. Therefore, the degrees of freedom during freeze-out annihilations are non-relativistic Majorana fermions, for which $M \gg T$, light Standard Model and dark particles (the dark Higgs and gauge boson). 

In order to make manifest the non-relativistic nature of the dark matter, one may write down a non-relativistic Lagrangian from the start. Moreover, non-relativistic particle
annihilations can be described by four-particle operators $\mathcal{O}_i$, arranged as an expansion in $1/M^2$. The prototype for such effective field theory (EFT) is the well-known non-relativistic QCD (NRQCD) \cite{Bodwin:1994jh}. The small parameter of the effective field theory is the average velocity $v \ll 1$ of the heavy particles, here the dark-matter Majorana fermions. In the EFT language, hard energy/momentum modes of order $M$ are \textit{integrated out} from the fundamental theory (\ref{Lag_UV_u1}). We write the low-energy Lagrangian explicitly in the next section~\ref{sec_U1_2}. The major benefit of the EFT formulation is to separate two classes of processes: those occurring
at the hard scale $M$, and those typical of the soft scales, either thermal or non-relativistic. Indeed, given the large energy release in the annihilation process, the typical distance scales are much smaller than those introduced by the thermal plasma, i.e.~$\Delta x \sim 1/M \ll 1/T$. 

Many scales remain still dynamical in the so-obtained low-energy theory: the thermal scales $\pi T$ and $g T$, and the non-relativistic scales $M\alpha$ and $M\alpha^2$.\footnote{$\pi T$ stands for the temperature scale where $\pi$ is a remnant of the Matsubara modes of thermal field theory, $gT$ identifies the scale of thermal masses with $g$ being a generic coupling constant, $Mv$ and $Mv^2$ the momentum and kinetic energy/binding energy of the heavy pairs. For coulombic and near-coulombic bound states $v \sim \alpha$ can be also used for the scales estimate.}   %Being the heavy Majorana fermions as part of a thermal bath, how does one can formulate annihilations in a medium? For sure 
At smaller energy scales, the heavy pairs can be sensitive to medium effects and a quantum statistical interpretation of pair annihilations is desirable. 
Since dark matter particles are slowly moving, repeated soft interactions can occur that are mediated by the dark Higgs and dark gauge boson. These interactions, that can modify the wave function of the annihilating dark matter pair, happen in a thermal bath. Hence, correlators should be evaluated within finite temperature field theory. It comes as the main strength of the approach exploited here~\cite{Kim:2016kxt,Kim:2016zyy,Biondini:2017ufr,Biondini:2018pwp} to recast the partition function of the annihilating pair as the thermal expectation value of the four-particle operators. This way one can dynamically account for the whole two-particle spectrum, both scattering and bound states properly weighted by the corresponding Boltzmann factor, and include near-threshold soft effects for which $T \sim M\alpha^2$. Bound states have an effect of order unity for such temperatures that is reflected in the Boltzmann factor of the annihilating pair (cfr. eq.~(\ref{Som_def})). For a more detailed and comprehensive discussion see refs.~\cite{Kim:2016kxt,Biondini:2017ufr}. 
 
In summary, we shall compute the thermally averaged annihilation cross section as $\langle \sigma v \rangle = \sum_i c_i \langle \mathcal{O}_i \rangle$, where a factorization of the heavy mass scale $M$ and the temperature is assumed, $M \gg T$. First, we have to derive the matching coefficients $c_i$ of non-relativistic four-particle operators $\mathcal{O}_i$ that create and annihilate dark matter pairs.  In a second stage, we shall compute the thermal average of the very same four-particle operators $\langle \mathcal{O}_i \rangle$ that amounts to solve a thermally modified Schr\"odinger equation for the dark matter pair with the thermal potentials of the mediators (see section~\ref{sec_U1_3} and \ref{sec_U1_4}). Finally, the extraction of the  corresponding spectral function comprises the information on the annihilating states in the statistical ensemble, i.e.\ scattering states and bound states.

\subsection{Non-relativistic Lagrangian}
\label{sec_U1_2}
In this section we outline the vertices between the heavy Majorana dark matter and the light degrees of freedom, namely the dark gauge boson $V^{\mu}$, the Goldstone boson $\varphi$ and the dark Higgs $s$ in the low-energy theory. This is the field theory that comprises energy modes with typical energies smaller than the dark matter mass. In addition we also write the four-particle operators describing the heavy Majorana fermion pair annihilations. We write the non-relativistic Majorana fermion as follows (we choose the standard parametrization of the Dirac matrices\footnote{We take the following assignment: $\gamma^0 = {\rm{diag}}(\mathbb{1}, -\mathbb{1})$, $\gamma^i=\left( \begin{array}{c c}
0 & \sigma^i \\
-\sigma^i & 0
\end{array} 
\right) $ and $\gamma^5=\left(  \begin{array}{c c}
0 & \mathbb{1} \\
\mathbb{1} & 0
\end{array} \right) $.})
\begin{equation}
\chi = \left( \begin{array}{c}
\psi \, e^{-iMt}
\\
-i \sigma_2 \psi^* e^{i Mt}
\end{array} \right) \, , \quad \bar{\chi}= \left( \psi^\dagger \, e^{i Mt} \, , \; -\psi^T i \sigma_2 e^{-iMt} \right) \, , 
\end{equation}
where the Grassmanian spinor $\psi$ has two components. Starting from the interaction Lagrangian after the U(1)$'$ symmetry breaking,
\begin{equation}
\mathcal{L}_{{\rm{int}}}=-\frac{g_\chi}{2} \bar{\chi}  \gamma^5 \gamma^\mu \chi V_{\mu} -\frac{y_\chi}{2 \sqrt{2}}\bar{\chi} \chi \, s +i \frac{y_\chi}{2 \sqrt{2}} \bar{\chi}  \gamma^5 \chi \, \varphi  + \cdots \, ,
\end{equation}
the terms which have no fast oscillations read
\begin{equation}
\mathcal{L}_{{\rm{int}}}^{{\rm{NR}}}= -  g_\chi \psi^\dagger_p (\bm{\sigma})_{pq} \psi_q \cdot \bm{V} - \frac{y_\chi}{ \sqrt{2}} \psi^\dagger_p \psi_p s + \cdots \, .
\label{NR_Lag_dark}
\end{equation}
The superscript stands for non-relativistic (NR) and $\sigma^i$ are the Pauli matrices. The Majorana fermion does not show any interaction with the temporal component of the gauge boson $V^{0}$, at variance with what happens in the case of heavy Dirac fermions interacting with gauge bosons via vector like currents, such as in the well-known Heavy Quark Effective Theory (HQEFT)\cite{Neubert:1993mb}, NRQCD\cite{Bodwin:1994jh} and potential NRQCD\cite{Pineda:1997bj,Brambilla:1999xf}. In our case, only the spatial components of the gauge field interact with the non-relativistic spinor.  An EFT approach for NR Majorana fermions has been previously introduced in ref.\cite{Biondini:2013xua}.  

Then we write down the absorpative Lagrangian that comprises the four-particle operators of the effective theory. Dealing with a Majorana fermion, there is only one operator at order $1/M^2$ which describes the dark matter annihilation\cite{Biondini:2018pwp}
\begin{equation}
\mathcal{L}_{{\rm{abs}}}= i c_1 \psi^\dagger_p \psi^\dagger_q \psi_q \psi_p \, ,
\label{Op_1_U1}
\end{equation} 
and we find the following matching coefficient\footnote{This result can be crosschecked with the cross sections given in \cite{Duerr:2016tmh} where more general expressions with finite masses for the particles in the final states are provided. We do not include suppressed operators of order $\mathcal{O}(1/M^4 )$ which correspond to $p$-wave annihilations.} %(s-wave annihilation, \textit{maybe it is worth looking at the p-wave operators?})
\begin{equation}
c_1= \frac{y_\chi^4 + 4 g_\chi^4}{64 \pi M^2} \, .
\label{c1_U1}
\end{equation}
According to the optical theorem, the imaginary part of the one-loop diagrams with four-particles external legs is equivalent to the matrix element squared of the annihilation processes of the type $\chi \chi \to a \, b$, where $a$ and $b$ are generic light degrees of freedom the heavy particles can annihilate into. Matching the four-point Green's function of the fundamental theory onto that of the low-energy theory fixes the coefficient given in eq.~(\ref{Op_1_U1}). This procedure is well established in the realm of non-relativistic effective field theories for QCD\cite{Bodwin:1994jh}.   
Since we are working with vanishing portal couplings, the possible final states are combinations of the real scalar, Goldstone boson and gauge boson referred to as \textit{dark terminators}\cite{Kahlhoefer:2015bea,
Duerr:2016tmh}. 

The annihilation cross section in the free case reads simply $
\langle \sigma v \rangle^{(0)}= 2 c_1 $.
%where the finite thermal mass correction affect both the equilibrium number density and the thermal expectation value of the four-particle operator. That is the reason why it does not appear in the expression of the annihilation cross section. 
For general orientation on the dark matter masses that provide the correct relic density, we anticipate some benchmark values to be $M \approx 0.5, \, 2, \, 5$ TeV for $y_\chi=0.5, \,1, \,1.5$ respectively and for $g_\chi=y_\chi/10$. 
%we show in figure \ref{fig:bench} (right panel) the corresponding benchmarks for different values of the Yukawa coupling, i.e.~$y_\chi = 0.5, 1, 1.5$ (in \cite{Duerr:2016tmh} values down to $y_\chi=0.2$ are considered).
%The term proportional to $y_\chi^4$ is due to the annihilation of the DM pair into the longitudinal components of the gauge bosons, being the annihilations into the real scalar $s$ is indeed p-wave suppressed \cite{Duerr:2016tmh}.

\subsection{Scalar and vector induced potentials}
\label{sec_U1_3}
The dark Higgs and the dark gauge boson can be exchanged between the dark matter pairs. 
If these particles are sufficiently lighter than the dark matter mass, they can induce sizeable effects on the scattering states, namely the Sommerfeld effect, and below threshold effects, i.e.~a bound state spectrum. Moreover, thermal effects can enter such dynamics and we include them in two respects. First, we use the scalar and gauge boson propagator in the Hard Thermal Loop (HTL) approximation \cite{Pisarski:1988vd,Braaten:1989mz,Frenkel:1989br,Taylor:1990ia}. In general the so-obtained propagators contain both a thermal mass and a finite thermal width that account for virtual and real scatterings with light degrees of freedom in the thermal plasma. Second, dark matter pairs interact in a statistical background and, therefore, their dynamics is properly described by correlators evaluated in a finite temperature field theory.  These can be expressed in terms of a spectral function at $T \neq 0$ that exhibits a smoothing between the bound state spectrum and the scattering states. 

Since the Majorana dark matter fermion couples to the spatial components of the vector boson (see eq.~(\ref{NR_Lag_dark})), the relevant self-energy is $\Pi^{ij}$. It is well known that in the static limit $\Pi^{ij}$ vanishes, namely there is no thermal mass nor imaginary part at one-loop order for the spatial gauge fields. 
However, the gauge field has a ``thermal mass'' through the temperature dependence of the dark-Higgs expectation value. The temperature dependent dark-Higgs expectation value reads (see appendix B for details) 
\begin{equation}
w^2_T=\frac{1}{\lambda_s} \left[ \frac{m_s^2}{2} - T^2 \left(  \frac{\lambda'}{3} +  g_\chi^2 \right)   \right]   \, ,
\label{U1_vev}
\end{equation}
from which we define $m_V^2 = 4 g_\chi^2 w^2_T$. When the temperature is such that $w_T \leq0$ the U(1)$'$ symmetry is restored and the mass of the dark gauge boson vanishes accordingly. %On the contrary of some other examples, such as the Inert Doublet Model, there is no residual pure thermal mass at high temperatures. This is because the coupling between the DM and the gauge boson involve spatial vectors.  

As far as the dark-Higgs propagator at finite temperature is concerned, we notice that no imaginary part arises in the HTL static limit. Only a finite thermal mass appears that is related to the expectation value already written in eq.~(\ref{U1_vev}). The dark-Higgs propagator reads, in the static limit and in the imaginary-time formalism
\begin{equation}
\lim_{\omega \to 0} i \langle s \, s \rangle _T (\omega,k) = \frac{1}{k^2 + 2 \lambda_s w^2_T}  =  \frac{1}{k^2 + m_s^2(T)} \, ,
\label{U1_sca_prop}
\end{equation}
where $k  \equiv |\bm{k}|$ and $m_s^2(T) \equiv 2 \lambda_s w^2_T $. 

We recall that by requiring small mediator masses, i.e.\ $M_\chi \gg m_V,m_s$, implies the condition $y_\chi \gg g_\chi, \lambda_s$. Hence, the interaction between the dark fermion and the dark scalar is parametrically more relevant than that involving the dark fermion and the gauge boson. 
Therefore, we focus on the interactions induced by fermion-scalar vertex in (\ref{NR_Lag_dark}) and we consider the corresponding diagrams in figure~\ref{fig:Higgs_potential_1}. The corresponding  thermal propagator is given in eq.~(\ref{U1_sca_prop}).
With the definition of an auxiliary potential function 
\begin{equation}
V_s(r) = y_\chi^2 \int_{\bm{k}} e^{i \bm{k} \cdot \bm{r}} \frac{1}{k^2+m_s^2(T)} \, ,
\label{VS_thermal_0}
\end{equation}
%= y_\chi^2 \left\lbrace \begin{array}{c}
%\frac{\exp{ \left( -m_s(T)r\right) } }{4 \pi r}  \, , \quad r>0
% \\
%\phantom{x} 
%\\
%-\frac{m_s(T)}{4 \pi}   \, , \quad r=0
%\end{array}  \right.  \, .
we write the dark-Higgs potential obtained from the diagrams in figure~\ref{fig:Higgs_potential_1} as
\begin{eqnarray}
\mathcal{V}_1 &=& - V_s(0) - V_s(r) 
\nonumber 
\\
&=& \alpha_y (m_s(T) - m_s) -\alpha_y \frac{e^{-m_s(T) r}}{r} \, , 
\label{VS_thermal}
\end{eqnarray}
where we defined $\alpha_y=y^2_\chi/(4 \pi)$.  We notice that the $r$-dependent part is attractive and the $r$-independent part provides an overall negative correction to the dark matter pair self-energy, given that $m_s(T)<m_s$. This can be traced back to a mass correction for a single dark matter particle.  Moreover, the $r$-independent part are linearly divergent, therefore the corresponding vacuum counterterms are
defined such that $\lim_{r \to \infty} \mathcal{V}_1(r) = 0$ at $T = 0$ \cite{Biondini:2017ufr}.  
%\begin{equation}
%M \to M + \Delta M_T = M + \frac{\alpha_y}{2} m_s(T) \, ,
%\end{equation}
%where the factor of one half appears because the potential refers to the pair, whereas here we write the contribution to one DM particle. \textit{Conversely to the case of DM-gauge bosons interactions, the mass correction can induce an effect that tends to decrease the annihilation rate}. 

In the next section we study the modification to the annihilation rate induced by the potential written in eq.~(\ref{VS_thermal}).  %As a first preliminary numerical output, we solve the Boltzmann equation with the free cross section for different $y_\chi$ in order to have an idea of the corresponding DM masses.   

\subsection{Plasma-modified Schr\"odinger equation and overclosure bound}
\label{sec_U1_4}
In order to compute the annihilation rate for a dark matter pair as part of a thermal bath, we use the formalism developed in refs.~\cite{Kim:2016zyy, Kim:2016kxt} and already applied for two dark matter models in refs.\cite{Biondini:2017ufr, Biondini:2018pwp}. At the core of the method is the extraction of a spectral function from the imaginary part of Green's functions
\begin{eqnarray}
&& \biggl[ 
   -\frac{\nabla_r^2}{M} + \mathcal{V}^{ }_i(r) - E'
 \biggr] G^{ }_i(E';\bm{r},\bm{r}')  =  
 N^{ }_i\, \delta^{(3)}(\bm{r}-\bm{r}') \, , 
 \\
 \phantom{s} \nonumber
 \\
&&\lim_{\bm{r},\bm{r}' \to \bm{0}} {\rm{Im}}  G^{ }_i(E';\bm{r},\bm{r}')
  =  \rho^{ }_i(E')  \, ,
\end{eqnarray} 
where the thermal potential is the one given in eq.~(\ref{VS_thermal}) and $N_i$ refers to the number of contractions of the four-particle operator. In this case there is only one operator with $N_1=2$, see eq.~(\ref{Op_1_U1}). In the potential induced by the dark Higgs there is no imaginary part within the approximation adopted in this work. However, we allow for a small imaginary part in the potential, i.e. $\mathcal{V}_1 -i \Gamma$, in order to extract the spectral function and we set it to $\Gamma \approx (10^{-6}$-$10^{-5})M$.\footnote{In practice the value of $\Gamma$ is is chosen to obtain numerical stability while keeping it as small as possible in order not to introduce fictitious effects. That said, it is possible to consider the decay width of the dark matter pair in the bound state. This choice has been made in the literature, see e.g. \cite{Garcia-Cely:2015khw}. However, it does not differ much from our choice since $\Gamma^{T=0} \approx \alpha_y^5 M /2$.}  

In this model we have to study a single spectral function corresponding to the annihilating Majorana fermion pair. Since there is no thermal width due to the Landau damping, the shape of the spectral function is rather insensitive to the value of the temperature. The dependence on the temperature enters the thermal dark-Higgs mass and  the couplings. The thermal mass $m_s(T)$ differs from the in-vacuum mass by up to 10\% depending the temperature and the model parameters. As far as $y_{\chi}$ is concerned, one has to evaluate it in a broad range of energy scales, namely $\mu \approx \pi T, e^{-\gamma_E}/r $ in the thermal potential (\ref{VS_thermal}) and at $\mu = 2 M$ in the matching coefficient of the hard annihilation in eq.~(\ref{c1_U1}). 
\begin{figure}[t!]
\centering
\includegraphics[scale=0.79]{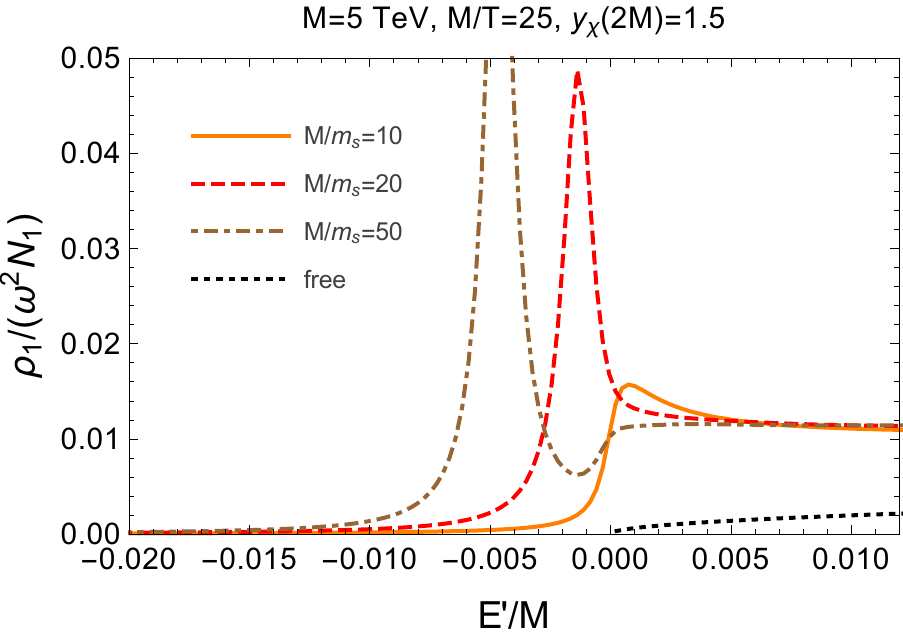}
\hspace{0.05 cm}
\includegraphics[scale=0.595]{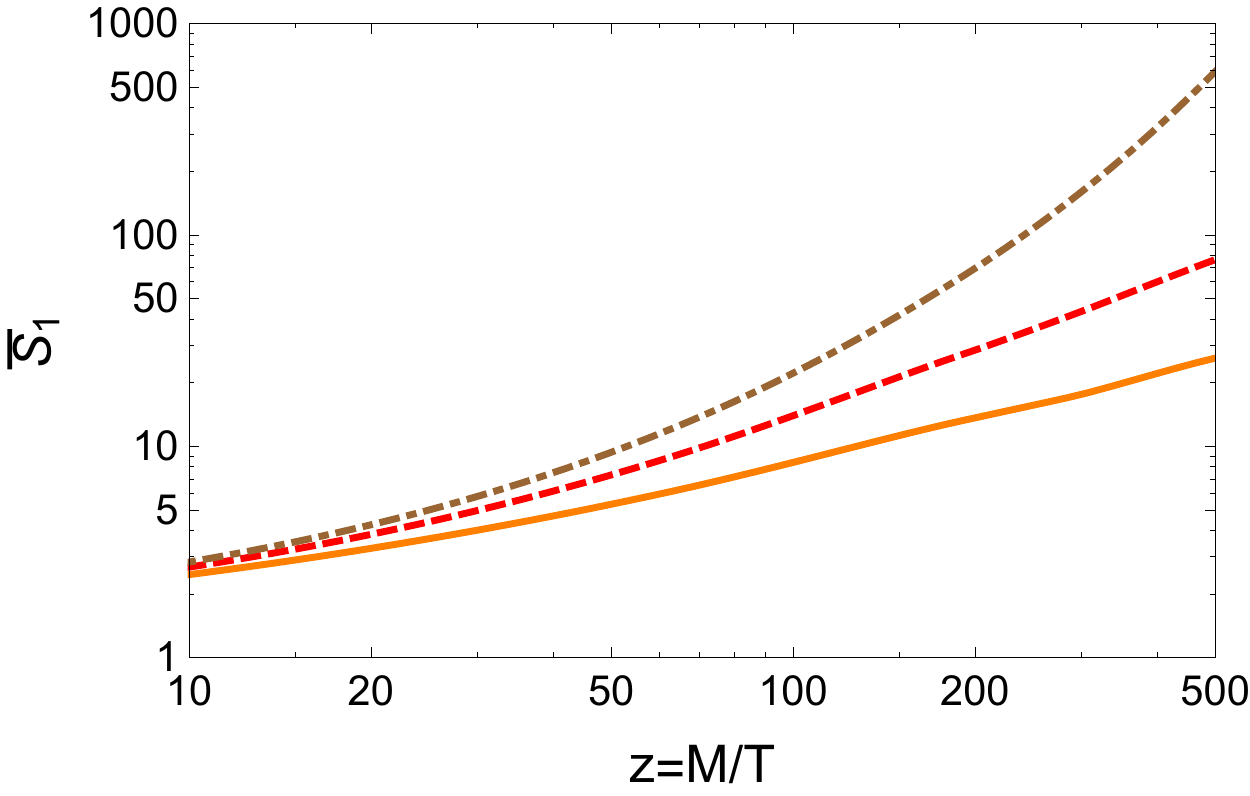}
\caption{\label{fig:spectral_y15}Left: Spectral function of the dark matter pair, here $\omega=2M+E'$. Three different ratios $M/m_s=10, \, 20, \, 50 $ are considered and $M=5$ TeV. Right: the thermally averaged Sommerfeld factor $\bar{S}_1$  for the same three mass ratios.}
\end{figure}
However, the main feature of the Laplace transform for the spectral function preserves the importance of bound states, if there are any (see eq.~(\ref{Som_def}) below). Whereas for $T$ larger than the binding energy the main contribution to the annihilation rate is given by the above threshold region, i.e.~Sommerfeld factors with appropriate thermal masses accounted for,  the bound state region dominates for $T \lesssim \alpha_y^2 M$.    
The generalized Sommerfeld factor is defined as follows\cite{Biondini:2017ufr,Biondini:2018pwp}
\begin{equation}
\bar{S}_1=\Bigl(\frac{4\pi}{MT} \Bigr)^{\frac{3}{2}}
 \int_{- \Lambda}^{\infty} 
 \! \frac{{\rm d}E'}{\pi} 
 \, e^{[ {\rm Re} \mathcal{V}^{ }_1(\infty)-E'] / {T}}
 \, \frac{ \rho^{ }_1(E') }{N^{ }_1}  \, ,
 \label{Som_def}
\end{equation}
where $\alpha_y^2 M \ll \Lambda \ll M$ is a cutoff restricting the average to the non-relativistic regime\cite{Kim:2016zyy,Kim:2016kxt}. 

In figure~\ref{fig:spectral_y15} we show the spectral function close to threshold for three different choices of the ratio between the dark matter fermion and the dark-Higgs masses $M/m_s$. The Yukawa coupling is chosen to be $y_\chi=1.5$ (it corresponds to $\alpha_y \approx 0.18$, pretty close to the largest value  considered in ref.\cite{Harz:2017dlj}, but smaller than the maximum value considered in ref.\cite{Duerr:2016tmh}, i.e.~$y_\chi=2$). A running Yukawa coupling has been included and it plays a role in a better estimation of the generalized Sommerfeld factors. Indeed, energy scales smaller than the hard annihilation scale are relevant in the Schr\"odinger equation, e.g.\ $\alpha_y M$ and $\pi T$. The Yukawa coupling $y_\chi$ decrease with the energy (see appendix A for details) at variance with what happens in QCD  and for the gluon exchange.
We look at a temperature around the freeze-out region, namely $M/T = 25$.  A bound state appears and is more prominent for smaller mediator masses, respectively $M/m_s=20$ and $M/m_s=50$ for the dashed-red and dot-dashed brown lines. The corresponding Sommerfeld factors, as defined in (\ref{Som_def}), are shown in the right panel of figure~\ref{fig:spectral_y15}. 

\begin{figure}[t!]
\centering
%\includegraphics[scale=0.8]{Spectral_Ex_y1_GammaOK}
%\hspace{0.05 cm}
\includegraphics[scale=0.56]{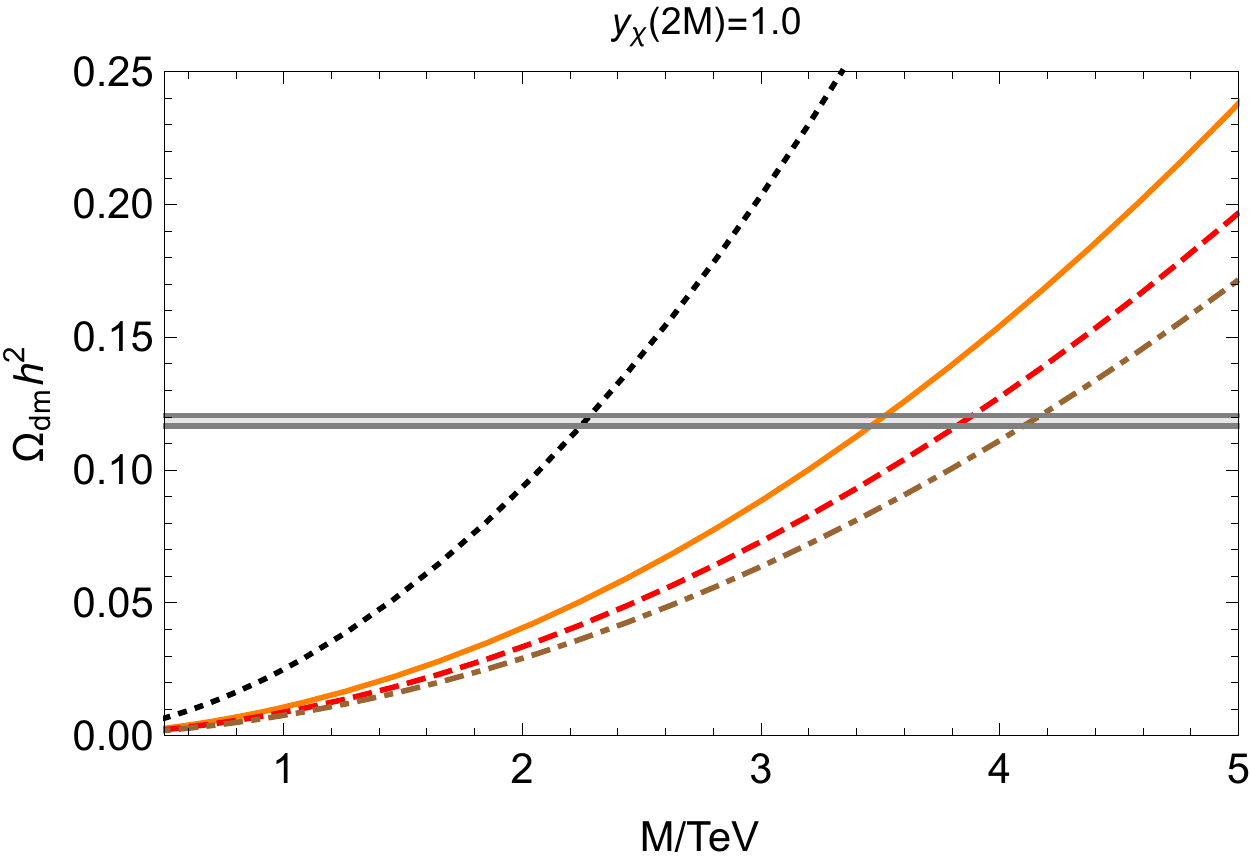}
\hspace{0.05 cm}
\includegraphics[scale=0.56]{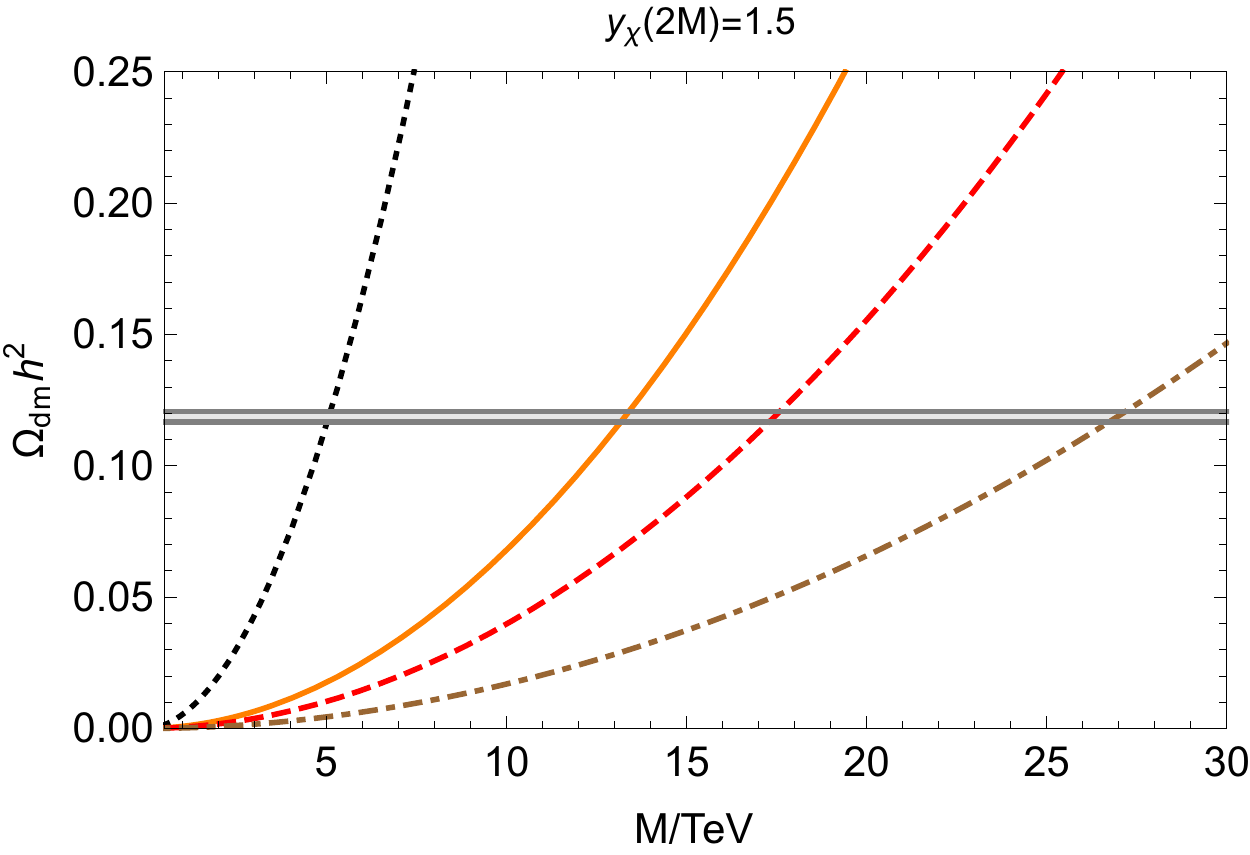}
\caption{\label{fig:overclosure_y1} Overclosure bounds for the case $y_\chi=1.0$ and $y_\chi=1.5$. The curves are obtained with the free cross section and with cross sections including the dark-Higgs exchange. The dark-matter dark-Higgs mass ratio is fixed according to the three choices $M/m_s=10, \, 20, \, 50$ and the color code is as in figure~\ref{fig:spectral_y15}.}
\end{figure}

Now we can proceed to the determination of the freeze-out abundance. Within a Boltzmann equation the  dark matter abundance evolves as\cite{Lee:1977ua,Gondolo:1990dk} (we label $n_\chi \equiv n$)
\begin{equation}
\dot{n}=-\langle \sigma v \rangle (n^2-n^2_{\hbox{\scriptsize eq}}) \, ,
\label{Bol_0}
\end{equation}
where $\dot{n}$ stands for the covariant time derivative in an expanding background. The thermally averaged annihilation cross section for the Majorana fermion pair reads 
\begin{equation}
\langle \sigma v \rangle = 2 c_1 \bar{S}_1 \, ,
\label{cross_def_T}
\end{equation}
where the generalized Sommerfeld factor $\bar{S}_1$ is extracted from the corresponding spectral function as in (\ref{Som_def})   and $c_1$ is from eq.~(\ref{c1_U1}). Then
we define the usual yield parameter $Y \equiv n/s$, where $s$ is the entropy density, and change variables
from time to $z \equiv M/T$. Therefore eq.~(\ref{Bol_0}) becomes
\begin{equation}
Y'(z)= - \langle \sigma v \rangle M m_{{\rm{Pl}}} \times \frac{c(T)}{\sqrt{24 \pi e(T)}} \times \left.  \frac{Y^2(z)-Y_{\hbox{\scriptsize eq}}^2(z)}{z^2} \right|_{T=M/z} \, ,
\end{equation}
where $m_{{\rm{Pl}}}$ is the Planck mass, $e$ is the energy density, and $c$ is the heat capacity, for which we
use values from ref. \cite{Laine:2015kra}.
In figure~\ref{fig:overclosure_y1} we show the overclosure bounds obtained with free cross sections and those accounting for the dark-Higgs exchange. On the left plot we set $y_\chi(2M)=1.0$ whereas $y_\chi(2M)=1.5$ in the right plot. In the latter case, the dark matter mass that reproduces $\Omega_{\hbox{\tiny DM}} h^2= 0.1186 \pm 0.0020$ is lifted from $M=5.1 \pm 0.1$ TeV to $M=(13.3,17.4,27.0) \pm 0.1$ TeV for the three ratios $M/m_s=10,20,50$ respectively. A smaller effect is observed for the first choice of the Yukawa coupling, where one finds an increase from free case $M=2.2 \pm 0.1$ TeV  to $M=(3.5,3.8,4.1) \pm 0.1$ TeV for the same $M/m_s$ values. The main reason for a smaller effect resides both in smaller Sommerfeld factors for the scattering states, together with less prominent bound states when passing from $y_\chi=1.5 \, (\alpha_y \approx 0.18)$ to $y_\chi=1 \, (\alpha_y \approx 0.08)$.

\begin{figure}[t!]
\centering
%\includegraphics[scale=0.8]{Spectral_Ex_y1_GammaOK}
%\hspace{0.05 cm}
\includegraphics[scale=0.7]{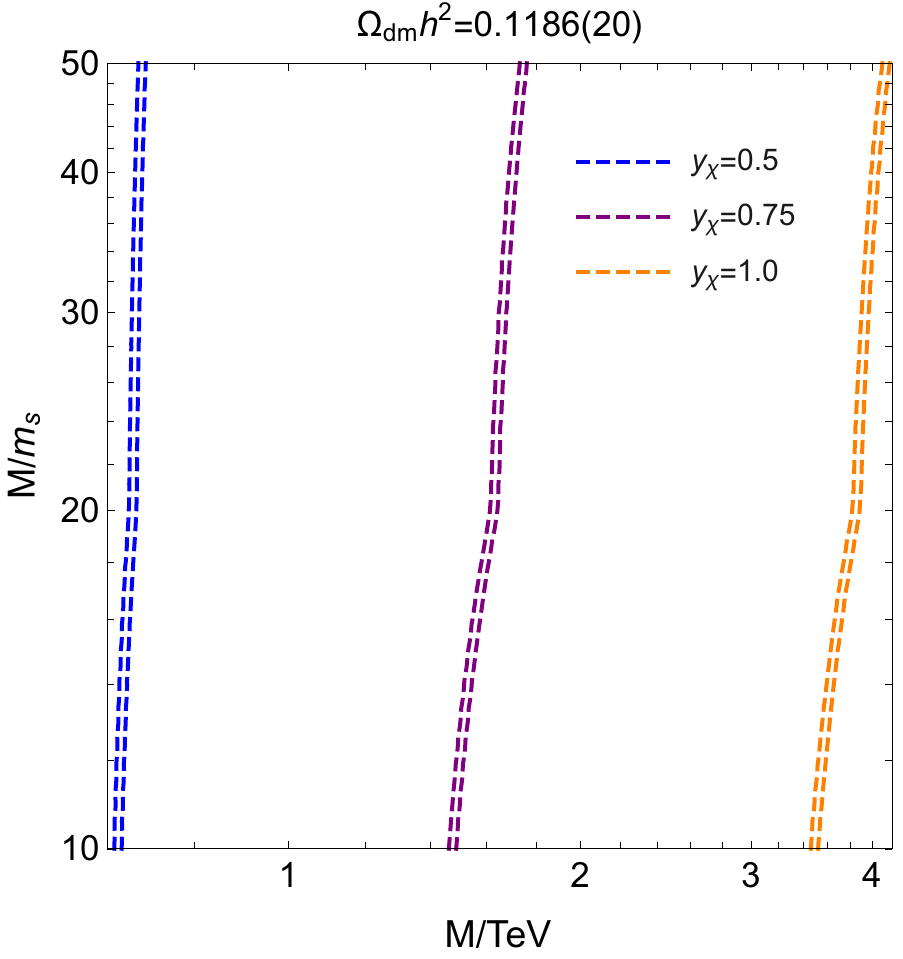}
\hspace{0.5 cm}
\includegraphics[scale=0.7]{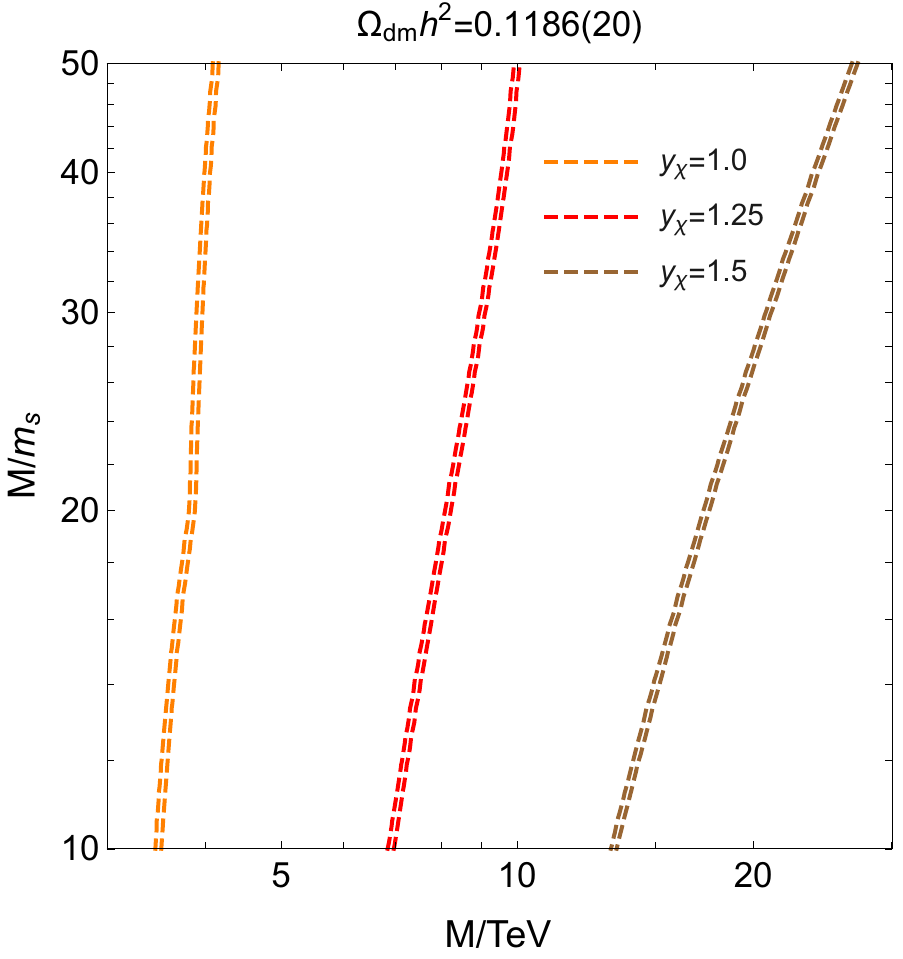}
\caption{\label{fig:overclosure_y_scan} The curves reproduce the correct dark matter relic abundance in the ($M,M/m_s$) plane. Different values of the Yukawa coupling $y_\chi$ are considered.}
\end{figure}

Finally, we show curves in the parameter space ($M,M/m_s$) for different values of $y_\chi$  that are compatible with the dark matter relic density in figure~\ref{fig:overclosure_y_scan}. For the smallest value $y_\chi=0.5$ ($\alpha_y \approx 0.02$) considered in this work, the increase due to the dark-Higgs exchange amounts at 20\% (25\%) for $M/m_s=10$ ($M/m_s=50$) lifting $M=0.55$ TeV to $M=0.66$ TeV ($M=0.70$ TeV). Therefore, according to the value of the Yukawa coupling, the corresponding effect on the overclosure bound ranges from an enhancement typical of weak interactions, as found in ref.\cite{Biondini:2017ufr}, up to larger effects observed in the case of strong interactions \cite{Biondini:2018pwp}.  
%{\color{red}In figure~\ref{fig:spectral_y1} we plot the spectral functions for a smaller value of the Yukawa coupling, namely $y_\chi=1$ ($\alpha_y \approx 0.08 $). Here we take  $M/m_s=10, 20, 30$. One may see that a smaller ratio $M/m_s$ is needed for a bound state to start forming with respect to the case with $y_\chi=1.5$. The shift of the threshold location decreases accordingly with smaller values of the dark-Higgs masses. DO WE PUT THIS FIGURE?}
\section{Other simplified models with Higgs-like exchange}
\label{sec_others}
In this section we address a different simplified model that comprises a trilinear vertex between a Higgs field and a dark matter pair. The model we have in mind comprises a Majorana dark matter particle coannihilating with a coloured scalar, the latter charged under QCD (see \cite{Garny:2015wea} for a review of the model). Besides the interactions with gluons and the corresponding potentials, additional effects induced by the Higgs exchange can appear.  We are not going to derive  the overclosure bounds as systematically as in the previous case, however we make contact with some of the results derived in section~\ref{sec_U1_main} when possible. 

We divide the discussion by following two different implementations of the interaction between the coloured scalar and the Higgs boson. First, we stick to the model we studied in \cite{Biondini:2018pwp}. In this case the interaction reads
\begin{equation}
\mathcal{L}_{{\rm{int}}}^{(1)} = -\lambda_3 \eta^\dagger \eta H^\dagger H + \dots \, ,
\label{OP_model_1}
\end{equation} 
where $\eta$ is the coloured scalar, $H$ the Standard Model Higgs doublet and $\lambda_3$ a scalar coupling. This Lagrangian 
leads to an interaction between the coloured scalar and the Standard Model Higgs that is suppressed by $v/M_\eta$  after the electroweak symmetry breaking. We want to assess whether relevant contributions to the generalized Sommerfeld factors can arise from this particular realization. 

Second, we start with the model as written in ref.\cite{Harz:2017dlj},
\begin{equation}
\mathcal{L}_{{\rm{int}}}^{(2)} = -g_h M_\eta \eta^\dagger \eta h + \dots \, ,
\label{OP_model_2}
\end{equation}  
where $h$ is taken to be a real scalar (possibly the Higgs boson) and there is no $1/M_\eta$ suppression after one expands $\eta$ in the non-relativistic modes. In this second option, the coupling between the real scalar and the coloured scalar is taken to be proportional to $M_\eta$ from the beginning (motivated by some SUSY arguments\cite{Harz:2012fz,Haber:1990aw}).
%We divide the discussion in the next two sections. 
\subsection{Case 1}
\label{sec_case1}
The impact of the gluon exchange for this model has been extensively studied \cite{Harz:2014gaa,Ellis:2014ipa,Ibarra:2015nca,Liew:2016hqo,Mitridate:2017izz,Keung:2017kot,Biondini:2018pwp}. The effect is particularly relevant when the mass splitting between the Majorana dark matter and the coloured scalar is small ($M_{\eta} \equiv M + \Delta M$ with $\Delta M \ll M$), so that the dark matter abundance is actually controlled by that of the $\eta$ particles, the latter experiencing strong interactions. Then Sommerfeld effects, decohering scatterings, bound state formation/dissociation have been included in the derivation of the freeze-out abundance. However, at temperatures $T \lesssim 160$ GeV, a trilinear coupling between the coloured scalar and the Higgs boson is established.  

After the electroweak symmetry breaking the trilinear vertex is given by the following non-relativistic Lagrangian 
\begin{equation}
\mathcal{L}_{{\rm{int}}}^{{\rm{NR}}}= -\frac{\lambda_3 v_T  }{2 M} \left( \varphi^\dagger \varphi + \phi^\dagger \phi \right) h \, ,
\label{NR_OP_1}
\end{equation} 
which is obtained from (\ref{OP_model_1})  when expanding $\eta=(\phi e^{-iM t}+\varphi^\dagger  e^{iM t} )/\sqrt{2 M}$ in terms of the non-relativistic fields $\phi$ and $\varphi$, where $\phi$($\varphi$) annihilates a particle (antiparticle). Then $h$ is the real scalar field corresponding to the Standard Model Higgs boson after the symmetry breaking. This vertex is suppressed by a factor $v/M$ with respect to the gluon induced one. The vertex is controlled by the temperature dependent Higgs expectation value, namely\cite{Kim:2016kxt}
\begin{equation}
v_T^2= \frac{1}{ \lambda} \left[ \frac{m_h^2}{2}- \frac{(g_1^2+3g_2^2 +8 \lambda+4 h_t^2)T^2}{16} \right]  \, .
\label{Higgs_vev_T}
\end{equation}
At temperatures larger than $T_c \approx 160$ GeV the Higgs mechanism \textit{melts} away and the trilinear vertex inducing the Higgs exchanges does as well.  The Higgs thermal mass squared reads
$m^2_h(T)=2 \lambda v_T^2$. 

The potential induced by the Higgs exchange can be calculated from the four-particle operators \cite{Biondini:2018pwp}
\begin{eqnarray}
 \mathcal{L}^{ }_{\hbox{\scriptsize abs}} & = & 
 i \, \Bigl\{ 
 c^{ }_1 \, 
 \psi^\dagger_p \psi^\dagger_q \psi^{ }_q \psi^{ }_p 
 + 
 c^{ }_2 \, 
 \bigl(
   \psi^\dagger_p \phi^\dagger_\alpha \psi^{ }_p \phi^{ }_\alpha +  
   \psi^\dagger_p \varphi^\dagger_\alpha \psi^{ }_p \varphi^{ }_\alpha   
 \bigr)
 \nonumber 
 \\
 & + & 
 c^{ }_3 \,  
\phi^\dagger_\alpha \varphi^\dagger_\alpha \varphi^{ }_\beta \phi^{ }_\beta
 + 
 c^{ }_4 \,  
\phi^\dagger_\alpha  \varphi^\dagger_\beta\,
 \varphi^{ }_\gamma \phi^{ }_\delta
 \, T^{a}_{\alpha\beta} T^{a}_{\gamma\delta}
  + 
 c^{ }_5 \, 
 \bigl( \phi^\dagger_\alpha \phi^\dagger_\beta
        \phi^{ }_\beta \phi^{ }_\alpha 
   +    \varphi^\dagger_\alpha \varphi^\dagger_\beta
        \varphi^{ }_\beta \varphi^{ }_\alpha 
 \bigr)
 \Bigr\} \, .
 \label{Lag_mod_2}
\end{eqnarray}
At variance with the $r$-dependent potentials induced by the gluon, that are different for each color representation, we obtain the same scalar contribution for all the operators
\begin{eqnarray}
&&\mathcal{V}_{2,h} = - V_h(0)/2 \, ,  \quad \mathcal{V}_{3,h} = \mathcal{V}_{4,h} = \mathcal{V}_{5,h}= -  \left[ V_h(0) + V_h(r)   \right] \, ,
\label{pot_Higgs_case1}
\end{eqnarray}
where we define the auxiliary thermal potential as  
\begin{equation}
V_h(r)= \left( \frac{\lambda_3 v_T}{2 M}\right)^2 \int_{\bm{k}} e^{i \bm{k} \cdot \bm{r}} \frac{1}{k^2+m_h^2(T)} \, .
\label{Higgs_potential}
\end{equation}
The exchange diagram gives an attractive potential for all the annihilation channels, on the contrary to what happens for the gluon exchange that induces a repulsive potential in the octet and $\eta$-$\eta$ operators (second, third and fourth operator in the second line of (\ref{Lag_mod_2})). We define an effective coupling 
\begin{equation}
\alpha_{\hbox{\scriptsize eff}} \equiv \frac{1}{4 \pi} \left( \frac{\lambda_3 v_T}{2 M}\right)^2 \, .
\label{alpha_eff_eq}
\end{equation} 
Exploiting the renormalization group equations (RGEs) derived in ref.\cite{Biondini:2018pwp}, we explore some possibilities according to different combinations for the model couplings (these are $\lambda_2$, $\lambda_3$ and $y$\cite{Garny:2015wea}). Moreover, we use $v_T$ as given in eq.~(\ref{Higgs_vev_T}). From figure~\ref{fig:compare_mod} (left panel), one may see that even in the case $\lambda_3=\pi$, which is a rather large value, we obtain at most $\alpha_{\hbox{\scriptsize eff}} \approx 0.01$. Based on our previous study involving such weak-interaction values for the coupling strength \cite{Biondini:2017ufr}, we conclude that the effect of the Higgs exchange  cannot compete with the gluon exchange in this realization of the model. Indeed, the corresponding generalized Sommerfeld factors induces an  increase of the overclosure bound of about 10\%, whereas QCD strong interactions give an increase of about 200\% for the same mass splitting  of the co-annihilating species (see figure 6 in ref.\cite{Biondini:2017ufr} and figure 3 in ref.\cite{Biondini:2018pwp} for $\Delta M = 5 \times 10^{-3}$).  
\subsection{Case 2}
\label{sec_case2}
The main reason for the smallness of the effective coupling in eq.~(\ref{alpha_eff_eq}) is the ratio $v/M$ originating from non-relativistic Lagrangian (\ref{NR_OP_1}). The simplified model considered in ref.\cite{Harz:2017dlj} is such that this suppression is absent and the coupling between the coloured scalar and the Higgs boson is taken to be in the range $\alpha_h \in [0.02,0.2]$, where $\alpha_h=g_h^2/(16 \pi)$\cite{Harz:2017dlj,Petraki:2015hla}. The largest and smallest value correspond almost to what we have considered in Section~\ref{sec_U1_main}, namely $y_\chi=1.5$ and $y_\chi=0.5$ that give $\alpha_y \approx 0.18$ and $\alpha_y \approx 0.02$. At this point the analysis carried out in section~\ref{sec_U1_main} can help in estimating the effect of the Higgs enhancement. 
\begin{figure}[t!]
\includegraphics[scale=0.62]{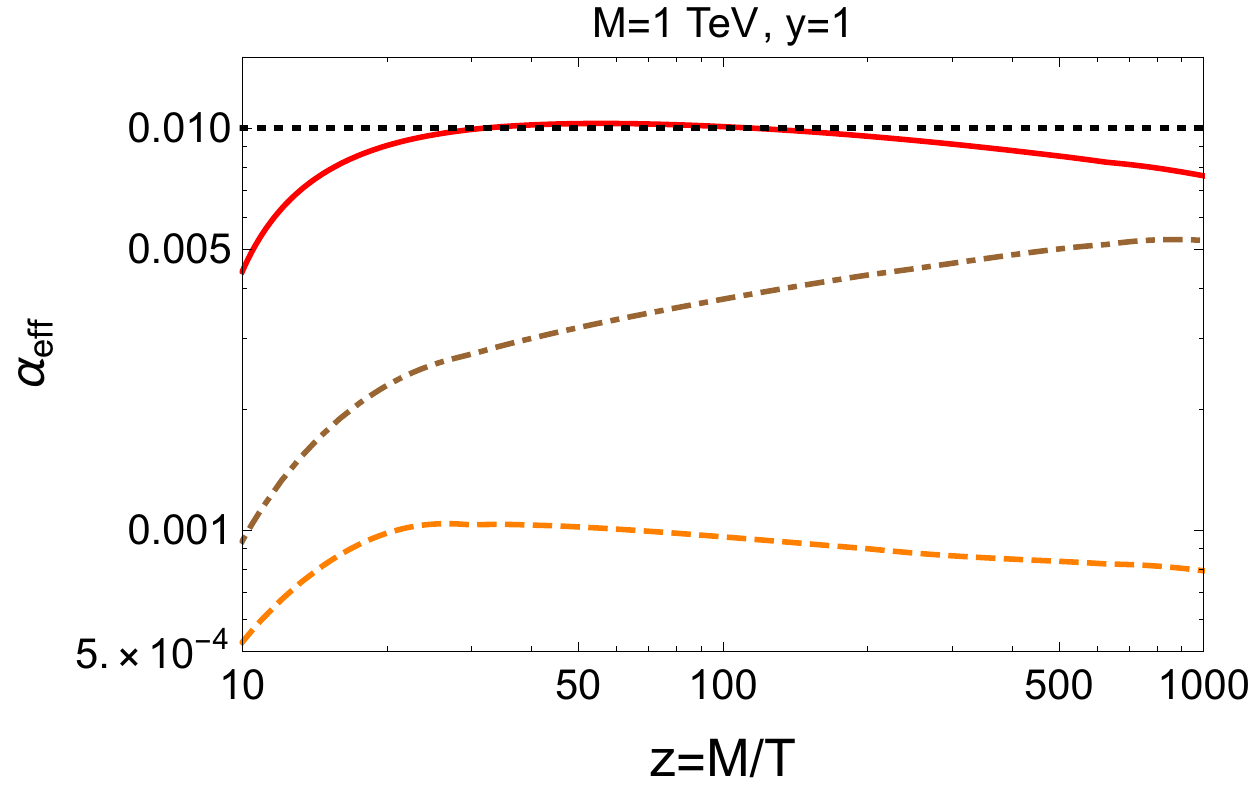}
\includegraphics[scale=0.83]{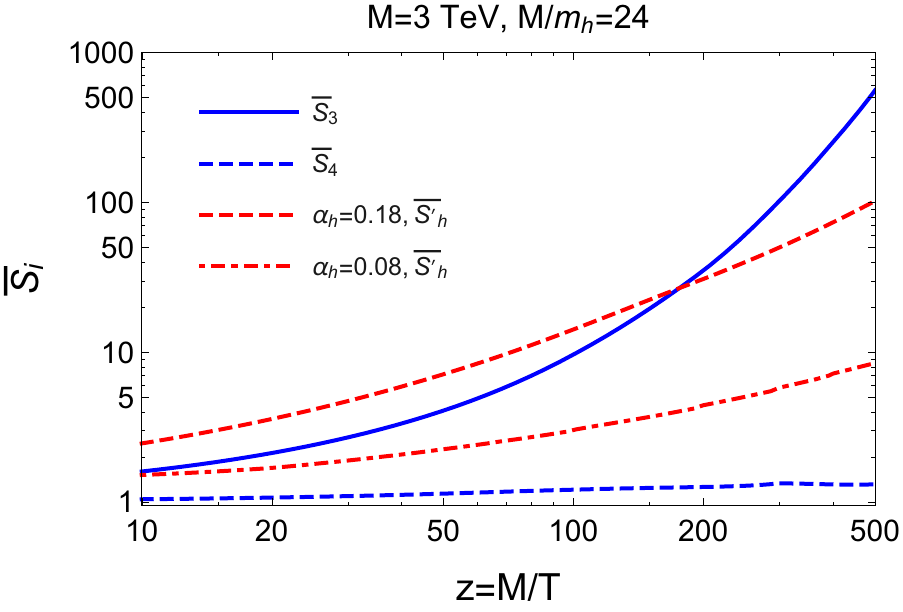}
\caption{\label{fig:compare_mod} Right plot: $\alpha_{\hbox{\scriptsize eff}}$ is given for $z=[10,10^3]$ for different combinations of the couplings, namely $\lambda_3=\pi, \, \lambda_2(2M)=0$ (solid red line), $\lambda_3(2M)=\pi/2, \, \lambda_2(2M)=0$ (dotted-dashed brown line), $\lambda_3(2M)=\pi/2, \, \lambda_2(2M)=\pi/2$ (dashed orange line). Left plot: generalized Sommerfeld factors due to gluon exchange (singlet and octet channel) and due to Higgs exchange.}
\end{figure} 

Let us start writing the cross section where both the gluon and Higgs exchange are included. We neglect the Majorana fermion $p$-wave suppressed operator and corresponding contribution to the cross section, and we then obtain 
\begin{equation}
\langle \sigma v \rangle  =\frac{ 4 c_2 N_c e^{-\Delta M_T/T} + N_c\left[  c_3 \bar{S}_3 + 
  c_4  \bar{S}_4 C_F +2 c_5 \bar{S}_5 (N_c+1) \right] e^{-2\Delta M_T/T}}{\left( 1+ N_c e^{-\Delta M_T/T} \right) }\, 
\label{cross_mod_2}
\end{equation}
where we split the thermally averaged Sommerfeld factors as 
\begin{equation}
\bar{S}_3 = \bar{S}_{3,g}+\bar{S}'_{h} \, , \quad \bar{S}_4 = \bar{S}_{4,g}+\bar{S}'_{h} \, , \quad \bar{S}_5 = \bar{S}_{5,g}+\bar{S}'_{h} \, .
\end{equation}
Here we write $\bar{S}'_{h}$ in order to signal that the Higgs-induced potentials have the same form as those given in eq.~(\ref{pot_Higgs_case1}), however the auxiliary potential reads in this case 
\begin{equation}
V'_h(r)= 4 \pi \alpha_h^2 \int_{\bm{k}} e^{i \bm{k} \cdot \bm{r}} \frac{1}{k^2+m_h^2(T)} \, .
\label{Higgs_potential_2}
\end{equation}
Once more we notice that the Higgs-induced generalized Sommerfeld factor is the same for all the operators and larger than unity. The thermal mass splitting entering eq.~(\ref{cross_mod_2}) has been derived in ref\cite{Biondini:2018pwp}.\footnote{At the level of this study we do not need the mass splitting, however for completeness let us mention that $\Delta M_T = \Delta M + g_s^2 C_F T^2/(12 M) -\alpha_s  C_F m_D(T)/2 +  \alpha_h (m_h(T)-m_h)/2$, where $\alpha_s=g_s^2/(4 \pi)$ and $\alpha_h=g_h^2/(16 \pi)$. The Higgs contribution is different from ref.\cite{Biondini:2018pwp} due to the different Lagrangian.} 

In figure~\ref{fig:compare_mod} we compare  the thermally averaged Sommerfeld factors induced by the gluon exchange (we take the results from ref.\cite{Biondini:2018pwp})  with those coming from the scalar exchange $\bar{S}'_h$, for a dark matter mass of $M=3$ TeV. Fixing the scalar mass to the Higgs boson mass $m_h=125$ GeV, we obtain the ratio $M/m_h=24$. We can borrow the results from the previous model, where we studied the generalized Sommerfeld factors for different dark-matter dark-Higgs mass ratio and for different Yukawa couplings. One can see that the attractive gluon exchange is already more important than the Higgs exchange for $\alpha_h = 0.08$ (that corresponds to $y_\chi \approx 1$), whereas the two processes provide a rather similar generalized Sommerfeld factor for  $\alpha_h = 0.18$.   
We note in passing that the Higgs exchange dominates over the gluon exchange in the octet channel (dashed-blue line in figure~\ref{fig:compare_mod}) as already noted in ref.\cite{Harz:2017dlj}. 

A comment is in order. The mediator mass that leads to a Yukawa potential has a different origin in the two cases. On one hand, the gluon mass is purely thermal, i.e.\ $m_D \approx g_s T$, and of order $10^2$ GeV for temperatures around the freeze-out for a dark matter mass at a TeV range. On the other hand, the Higgs mass is $m_h(T)=\sqrt{2 \lambda}v_T$ with $v_T$ from eq.~(\ref{Higgs_vev_T}). Here the thermal contributions play a little role around the freeze-out temperature, making the in-vacuum mass the relevant mass scale of the exchanged particle. 
%In this case, we see from figure~\ref{fig:spectral_y15} that a bound state start to form for $M/m_s=10$, that could represent a benchmark for a Higgs boson mass $m_h=125$ GeV and a dark matter mass $M \approx 1$ TeV.   In summary we expect that the Sommerfeld factor induced by the Higgs exchange provides a large contribution here, that can potentially be of the same order of the gluon induced one. 
 
\section{Conclusions and discussion}
\label{sec_concl}
In this paper we have studied the impact of a scalar exchange on the dark matter relic density. In order to quantify such an effect, we considered a simplified model with a Majorana dark matter fermion charged under a new U(1)$'$ gauge group. The dark sector is made of a dark gauge boson and a dark Higgs boson in addition to the Majorana fermion. The dark vector and scalar are the model mediators, and can possibly interact with Standard Model particles. We restrict our study to the case of vanishing portal couplings ($\kappa=\lambda_{hs}=0$ in eq.~(\ref{Lag_UV_u1})) and we assume light mediators, $M \gg m_s,m_V$. The latter assumption implies that the coupling between the dark matter and the dark scalar is larger than that between the dark matter and the gauge boson. 

Profiting from an effecting field theory framework, we derived the non-relativistic Lagrangian that describes the interaction between the heavy dark matter fermion and the light degrees of freedom. The impact of the dark-Higgs exchange is taken into account by solving a thermally modified Schr\"odinger equation and extracting the generalized Sommerfeld factors from a spectral function. Then the Boltzmann equation is solved with the corresponding annihilation cross section. We scan over the Yukawa coupling $y_\chi \in [0.5,1.5]$ and for the dark-Higgs mass $m_s \in [M/50,M/10]$. Going to lighter scalar masses, a larger impact of the scalar exchange is observed and bound states appear for sufficiently large values of the Yukawa couplings. Our results complement previous works where the scalar exchange has been considered and we add a possible treatment of bound-state effects. As already observed in the case of the gauge boson exchange (weak gauge bosons and gluons), we find that the dark matter mass reproducing the observed relic abundance is shifted to larger values with respect to the tree level one. However, this enhancement depends crucially  on the parameters of the model at hand (see figures~\ref{fig:overclosure_y1} and \ref{fig:overclosure_y_scan}), namely the coupling $y_\chi$ and the scalar mass $m_s$. The generalized Sommerfeld factors obtained in this model are effective down to low temperatures because there is no suppression given by  mass splittings with any coannihilating specie, i.e. $e^{-\Delta/T}$. These results are collected in section~\ref{sec_U1_main}.

In addition, we compared the generalized Sommerfeld factors coming from the interactions with gluons and a Standard Model Higgs boson for a different simplified model in section~\ref{sec_others}. Here, the impact of the scalar exchange depends on how the interaction between the coloured scalar and the real scalar is implemented (see section~\ref{sec_case1} and \ref{sec_case2}). We find that the Higgs exchange can induce an effect as large as the gluon exchange and lead to bound-states formation.
   
Finally, let us remark that the scalar exchange can affect the overcloure bounds significantly and should be then included in the relic density calculation. For the simplified model with a Majorana dark matter and a dark Higgs,  the dark matter mass is lifted from $(0.55, 2.2, 5.1)$ TeV to $(0.70, 4.1, 27.0)$ TeV respectively for three benchmark values $y_\chi=(0.5, 1, 1.5)$ considered in this work. The parameter space that reproduces the observed dark matter abundance is rather modified  and the overclosure bound is pushed to larger masses. In light of these results, it seems worth exploring the impact on direct and indirect searches for the very same model in order to better assess the reach of present and upcoming experiments (such as XENON1T\cite{Aprile:2017aty}, DARWIN\cite{Aalbers:2016jon} and CTA\cite{Carr:2015hta}) in the medium/high mass range.  %Finally, we notice that the generalized Sommerfeld factors lead to a larger enhancement in the U()$'$ model with respect to other cases where an in-vacuum mass splitting controls the large Sommerfeld factors at small temperatures (see e.g.~\cite{Biondini:2018pwp}).   
%Finally, the results presented here can be ameliorated in different respects. First of all the portal couplings can be included, especially the one that mixes the dark Higgs boson with the Standard Model one. This brings to a coupled evolution of the expectation values of the dark and visible sector perhaps setting some constraints on the possible choice for the scalar coupling $\lambda_s$ and the dark Higgs mass $m_s$. This can have consequences on the values for $M/m_s$ used in this work. Moreover, since a non-vanishing $\lambda_{hs}$ determines a scalar mixing, the dark matter fermion would couple with an admixture of the dark scalar and the Higgs boson, namely two scalar particles and not only one. Second, we used a HTL propagator in the static limit for the dark scalar that has a vanishing imaginary part (only virtual corrections appear as thermal masses). Going beyond this limit would allow to include real scattering with light particles in the plasma so to account for bound-state dissociation processes.  
 \section*{Acknowledgements}
This work was supported by the Swiss National Science Foundation (SNF) under grant
200020-168988. The author thanks Mikko Laine for valuable discussions and comments on the manuscript. 

\appendix
\numberwithin{equation}{section}
\section{Renormalization Group Equations}
\label{App_A}
In this section we present the results for the running couplings relevant for the U(1)$'$ model. In particular, we need the coupling $y_{\chi}$ for a broad range of energies: from $E \sim 2M$, the typical energy scale of the hard annihilations, to  $E \sim \pi T , e^{-\gamma_E}/r$, where the very same coupling is evaluated in the thermal potential. RGEs for a similar model have been derived in ref.\cite{Kajantie:1997hn}. 

We use dimensional regularization in the $\overline{{\rm{MS}}}$ with $D=4-2\varepsilon$ and compute the diagrams in the Feynman gauge. %First, we write the counterterms for the wave function and the vertices renormalization (we show the result for the Majorana fermion-gauge boson vertex\footnote{we checked explicitly gauge invariance by computing also the dark Higgs-gauge boson vertex counterterm.}):
%\begin{eqnarray}
%&&\delta_\chi = - \frac{\mu^{-2\varepsilon}}{(4 \pi)^2 \varepsilon} \left( \frac{y_\chi^2}{2} + g_\chi^2 \right) \, , 
%\\
%&&\delta_S= -   \frac{\mu^{-2\varepsilon}}{(4 \pi)^2 \varepsilon} \left( \frac{y_\chi^2}{2} - 8 g_\chi^2 \right) \, ,
%\\
%&&\delta_V = -   \frac{\mu^{-2\varepsilon}}{(4 \pi)^2 \varepsilon}  \left( \frac{2 N_F + 4 N_s + 4 N_f (q_f/q_\chi)^2}{3} g_\chi^2 \right) \, ,
%\\
%&&\delta_{g_\chi}= -\frac{\mu^{-2\varepsilon}}{(4 \pi)^2 \varepsilon}  \left(\frac{y_\chi^2}{2} +  g_\chi^2  \right) \, ,
%\\
%&&\delta_{y_\chi}= -\frac{\mu^{-2\varepsilon}}{(4 \pi)^2 \varepsilon}  \left( - y_\chi^2 + 4  g_\chi^2 \right) \, .
%\end{eqnarray}
%In the vector boson self-energy counterterm we explicitly write the contribution from the number of fermions and the scalars. 
%We use the following equations to determine the running
%&&\beta(y)=\mu \frac{\partial }{\partial \mu} \left[  -y \delta_y + \frac{y}{2} \left( 2 \delta_\chi + \delta_S  \right)  \right] \, , 
%\\
%&&\beta(g)=\mu \frac{\partial }{\partial \mu} \left[  -g \delta_g + \frac{g}{2} \left( 2 \delta_\chi + \delta_V \right)  \right] \, ,
%\end{eqnarray}
The RGEs  read at one loop for the relevant couplings
%\mu^{-2 \varepsilon}
\begin{eqnarray}
&&\mu \frac{d g_\chi^2}{d \mu} = \frac{1}{8 \pi^2} \left( \frac{2 N_F + 4 N_s +  4 N_f (q_f/q_\chi)^2}{3} g_\chi^2 \right) \, ,
\\
&&\mu \frac{d y_\chi^2}{d \mu} = \frac{1}{8 \pi^2} \left( \frac{7}{2} y_\chi^4 - 10 y_\chi^2g_\chi^2 \right) \, ,
\\
&&\mu \frac{d \lambda_s}{d \mu} = \frac{1}{8 \pi^2} \left( 48 g_\chi^2 - 24 g_\chi^2 \lambda_s + 10 \lambda_s^2 - y^4 + y_\chi^2  \lambda_s  \right) \, .
\end{eqnarray}
In the running for the $g_\chi$, that is fixed by the wave-function renormalization of the vector boson, we show the different contributions explicitly (dark matter fermion, dark scalar and Standard Model fermion). In our case we have $N_F=1$ and $N_s=1$ and we set $q_f=0$ for all the $N_f$ Standard Model fermions. In the numerical evaluation we simply impose $g_\chi=y_\chi /10$, in order  to satisfy the relation $y_\chi \gg g_\chi$.   
%(\textit{I still have to compute the renormalization of $\lambda'$ and cross check carefully with the equations given in \cite{Kajantie:1997hn} for a very similar model.})
\section{Dark-Higgs self-energy}
\label{App_B}
\begin{figure}
\centering
\includegraphics[scale=0.42]{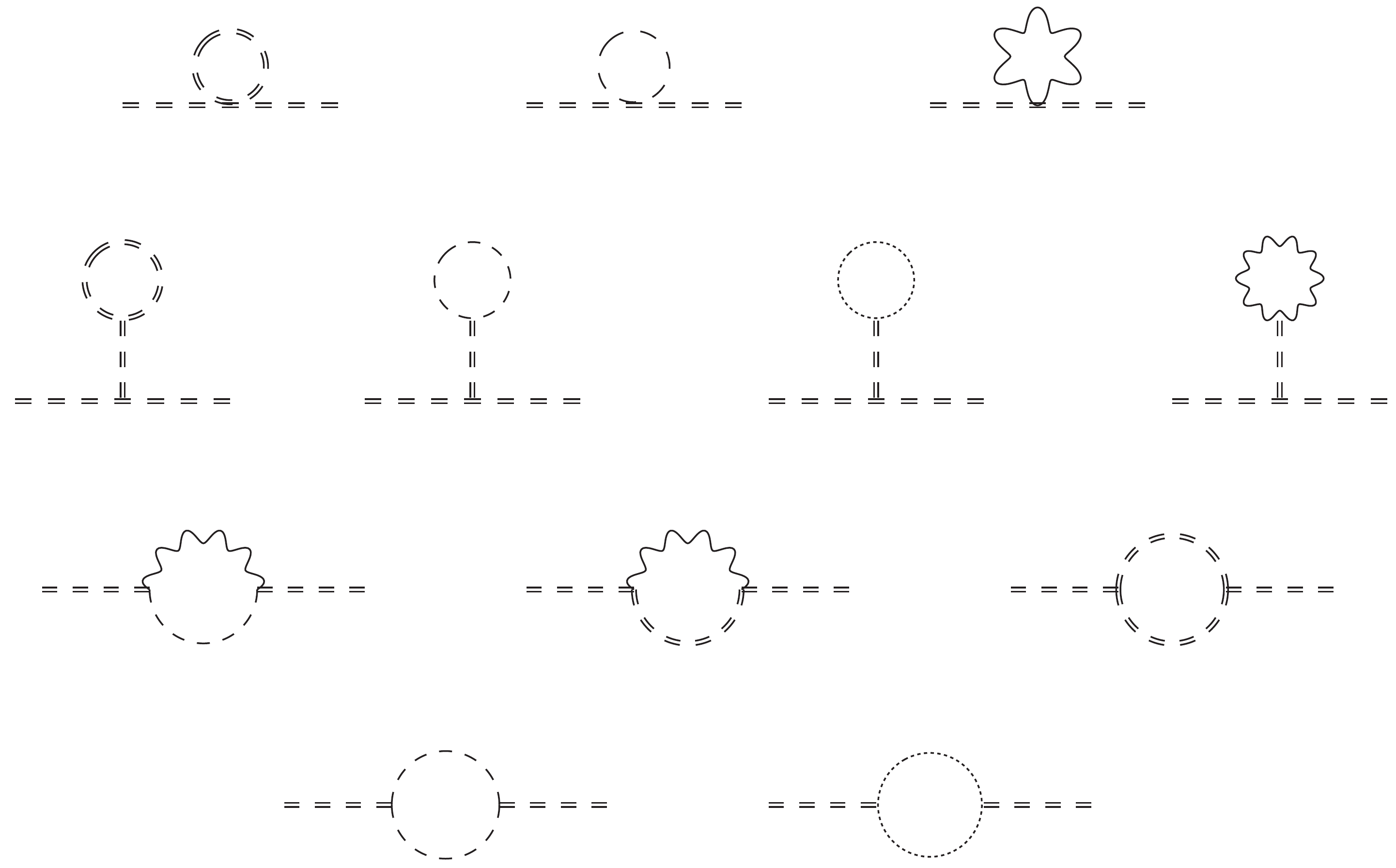}
\caption{\label{fig:appendix_2}One-loop diagrams for the dark-Higgs self-energy in Feynman gauge. Double-dashed lines stand for the dark Higgs, dashed lines for the Goldstone boson, wiggly lines for the gauge boson and dotted lines for the ghost field.}
\end{figure}

The dark scalar self-energy at finite temperature has been used in the body of the paper. The thermal self-energies for a dark gauge boson and dark scalar in a model very similar to the one we studied here can be found in ref.\cite{Kim:2016zyy}. As far as the dark-Higgs self-energy is concerned, the diagrams are shown in figure~\ref{fig:appendix_2} for the Feynman gauge (ghosts and Goldstone bosons are included in the diagrams). Our result agrees with that in ref.\cite{Kim:2016zyy}, upon the change $e' \to 2 e' \equiv 2 g_\chi$. The self-energy in the imaginary time formalism reads in the HTL limit 
\begin{equation}
\Pi_s = \left[ 8 g_\chi^2 (D-1) + 8 \lambda_s \right] \frac{T^2}{12} \, .
\end{equation}
Then the finite temperature dark-Higgs expectation value is 
\begin{equation}
w_T^2= \frac{1}{\lambda_s} \left[ \frac{m^2_s}{2} -\frac{T^2}{12} (24 g_\chi^2 + 4 \lambda_s) \right] \, ,
\end{equation} 
that gives eq.~(\ref{U1_vev}) and where $m_s$ is the $T=0$ scalar mass. 
\bibliographystyle{hieeetr}
\bibliography{higgsDM.bib}

\end{document}